\documentclass[final,3p,times, 11pt]{elsarticle}
\usepackage{graphicx, color}
\usepackage{bm}

\usepackage{algorithm}
\usepackage{algpseudocode}
\algnewcommand{\Initialize}[1]{%
  \State \textbf{Initialize:}
  \State \hspace*{\algorithmicindent}\parbox[t]{0.8\linewidth}{\raggedright #1}
}

\usepackage{amsmath}

\journal{Aerospace Science and Technology}

\begin{document}

\begin{frontmatter}

\title{Variational Mode Decomposition--Based Nonstationary Coherent Structure Analysis for Spatiotemporal Data}

\author[label1]{Yuya Ohmichi\corref{mycorrespondingauthor}}
\address[label1]{Aviation Technology Directorate, Japan Aerospace Exploration Agency, 7-44-1 Jindaiji-Higashi, Chofu, Tokyo 182-8522, Japan}
\cortext[mycorrespondingauthor]{Corresponding author}

\ead{ohmichi.yuya@jaxa.jp}

\begin{abstract}
The conventional modal analysis techniques face difficulties in handling nonstationary phenomena, such as transient, nonperiodic, or intermittent phenomena.
This paper presents a variational mode decomposition--based nonstationary coherent structure (VMD-NCS) analysis that enables the extraction and analysis of coherent structures in the case of nonstationary phenomena from high-dimensional spatiotemporal data.
The VMD-NCS analysis decomposes the input spatiotemporal data into intrinsic coherent structures (ICSs) that represent nonstationary spatiotemporal patterns and exhibit coherence in both spatial and temporal directions.
Furthermore, unlike many conventional modal analysis techniques, the proposed method accounts for the temporal changes in the spatial distribution with time.
The performance of the VMD-NCS analysis was validated based on the transient growth phenomena in the flow around a cylinder.
It was confirmed that the temporal changes in the spatial distribution, depicting the transient growth of vortex shedding where fluctuations arising in the far-wake region gradually approach the near-wake region, were represented as a single ICS.
Furthermore, in the analysis of the quasi-periodic flow field around a pitching airfoil, the temporal changes in the spatial distribution and the amplitude of vortex shedding behind the airfoil, influenced by the pitching motion of the airfoil, were captured as a single ICS.
Additionally, the impact of two parameters that control the number of ICSs ($K$) and the penalty factor related to the temporal coherence ($\alpha$), was investigated.
The results revealed that $K$ has a significant impact on the VMD-NCS analysis results.
In the case of a relatively high $K$, the VMD-NCS analysis tends to extract more periodic spatiotemporal patterns resembling the results of dynamic mode decomposition. In the case of a small $K$, it tends to extract more nonstationary spatiotemporal patterns.
\end{abstract}

\end{frontmatter}

\section{Introduction}
Generally, many physical phenomena generate complex spatiotemporal structures owing to their nonlinearity. In recent years, techniques for extracting spatiotemporal patterns from high-dimensional time-series data obtained through computational fluid dynamics or experiments have been actively researched in the fluid dynamics field. These techniques aim to elucidate the physically significant fluid structures and their temporal behaviors within complex flow fields. Additionally, they are employed to model the flow field for prediction and control \cite{Taira2017,Rowley2017,Taira2020,Unnikrishnan2023}.
The most common analytical methods are proper orthogonal decomposition (POD) \cite{Lumley1967, Berkooz1993} and dynamic mode decomposition (DMD) \cite{Schmid2010}. POD identifies the dominant spatiotemporal patterns by computing orthogonal bases that minimize the reconstruction error when reconstructing the original data. DMD represents the flow field's temporal evolution from one time to the next using a linear operator, thereby extracting modes as eigenvalues and eigenvectors of the operator.
Due to the linear-operator representation, each mode exhibits a linear growth in periodic oscillation amplitudes, simplifying the physical interpretation of the obtained modes.
These techniques have found extensive applications in the extraction of spatiotemporal patterns from complex flow fields, such as jets \cite{Rowley2009,Jovanovic2014,Lim2020}, wakes \cite{Muld2012, Ohmichi2019, Placco2023}, and flows involving shockwaves \cite{Grilli2012, Ohmichi2018, Giannelis2020}.
Moreover, substantial research efforts have been devoted to developing advanced methodologies, encompassing algorithms tailored for large and streaming datasets \cite{Hemati2014, Ohmichi2017c, Amor2023}, noise-robust algorithms \cite{Dawson2016,Hemati2017,Nonomura2019, Sashidha2022,Ohmichi2022ttls}, data-reconstruction methods from noisy \cite{Pastuhoff2013, Sugioka2019, Ohmichi2022} or incomplete \cite{Everson95, Nekkanti2023} data, and stability analysis methods \cite{Ranjan2020,Stahl2023}.

However, existing modal analysis techniques encounter difficulties in handling transient, nonperiodic, or intermittent phenomena (nonstationary phenomena).
For instance, the DMD method represents each mode's temporal behavior as a linear growth and periodic oscillation. Thus, it is unable to capture temporal variations in amplitude and frequency inherent in nonstationary phenomena.
The POD method is frequently limited by the coexistence of various phenomena within a single mode, complicating its physical interpretation. Additionally, most existing modal-decomposition methods rely on input data representations of the following form \cite{Noack2016}:
\begin{equation}
    f(\bm x, t) = \sum_{i=1}^{N} {w_i(t) \phi_i(\bm x)} + \epsilon(\bm x, t) 
    \label{eq:mode_decomp}
\end{equation} 
Here, the input data are represented as a superposition of $N$ space-dependent modes, $\phi_i$.
The coefficients, $w_i$, are expansion coefficients that vary with time.
$\epsilon$ denotes the expansion residual.
Equation (\ref{eq:mode_decomp}) indicates that the spatial distribution of mode $\phi_i$ remains unchanged with time.
However, many nonstationary phenomena, such as the transient growth of a cylinder wake, involve spatial structures that vary with time.
Therefore, to extract such phenomena as a single mode, a modal expansion using `flexible' modes capable of changing spatial structures is required, as noted by Noack \cite{Noack2016}.

In the field of signal processing, variational mode decomposition (VMD) \cite{Dragomiretskiy2014a} and its multivariate counterpart, multivariate VMD (MVMD) \cite{Rehman2019}, have recently found applications as analytical tools for addressing nonstationary time-varying data.
VMD tackles optimization problems by incorporating penalty terms related to the frequency bandwidth of modes, thereby extracting the coherent fluctuation patterns (modes) within the input signal.
Unlike DMD, VMD does not assume single periodic variations and can represent nonlinear changes in amplitude.
Each mode exhibits a distinct frequency spectrum with different center frequencies.
Additionally, by extracting modes with narrow frequency bandwidths, it becomes feasible to isolate similar frequency variations as one coherent phenomenon, enhancing the ease of deriving physical interpretations.
However, directly applying VMD algorithms to high-dimensional time-series data, such as fluid data, is challenging.

Only a few studies have attempted to develop modal analysis techniques for nonstationary fluid phenomena. The study by Liao et al. \cite{Liao2023} is the sole study focusing on an approach based on VMD.
They proposed reduced-order VMD (RVMD) inspired by the Hilbert–Huang transform and VMD.
However, RVMD is a modal-decomposition method that is represented as Equation (\ref{eq:mode_decomp}), which cannot capture the temporal evolution of spatial distributions of nonstationary phenomena as a single mode.
In the approach based on DMD, called multi-resolution DMD (mrDMD) \cite{Kutz2016}, different dynamic modes are extracted for each temporal domain of data based on the repetitive removal of low-frequency components and data segmentation.
The frequency-time analysis using the spectral POD is also based on data segmentation \cite{Nekkanti2021}.
The conditional space--time POD method \cite{Schmidt2019} extracts dominant spatiotemporal patterns from a set of event realizations.
The modes of the conditional space-time POD exhibit coherent structures in both spatial and temporal directions.
Recently, Stahl et al. \cite{Stahl2023} reported that convective instability can be analyzed by applying DMD to conditional space-time POD modes.

In this study, a variational mode decomposition--based nonstationary coherent structure (VMD-NCS) analysis method was developed for extracting and analyzing spatiotemporal patterns of nonstationary phenomena, and its effectiveness was validated.
The proposed method combines dimensionality reduction via POD and MVMD to achieve modal expansion using `flexible' modes for high-dimensional spatiotemporal data.
The proposed method was applied to transient phenomena in the wake region of a square cylinder and the flow around a pitching airfoil to demonstrate its capability in capturing spatiotemporal patterns in nonstationary phenomena.

\section{Methodology}
\subsection{Variational Mode Decomposition}
In VMD, a single-variable time series, $y(t)$, is decomposed into multiple time-series patterns, called VMD modes, $u_k(t),~(k=1,2,\cdots,K)$.
\begin{equation}
    y(t) =  \sum_{k=1}^{K} u_k(t),
\end{equation}
where $K$ is the number of modes.
VMD is a data-driven modal-decomposition method, and each mode, $u_k(t)$, is computed from the input signal, $y(t)$.
The purpose of VMD is to represent the input signal as a sum of modes $u_k(t)$, each with narrow-band frequency components.
To achieve this, the modes are determined such that the sum of the bandwidths of all modes is minimized.
To estimate the bandwidth of each mode, the following approach was proposed in the orignial VMD algorithm:
1) Compute the analytic signal for each mode via the Hilbert transform, resulting in a single-sided frequency spectrum.
2) Shift the frequency spectrum of each mode to the baseband by multiplying it with a complex exponential at the respective estimated center frequency.
3) Estimate the bandwidth by taking the squared L2 norm of the gradient of the harmonically mixed signal obtained in the previous step.
This formulation leads to the following constrained variational problem, aiming to minimize the bandwidth of the frequency distribution of each VMD mode and the reconstruction error:
\begin{equation}
\begin{aligned}
& \underset{\{u_k\},~\{\omega_k\}} {\text{minimize}} &&\sum_{k=1}^{K} \left\lVert \frac{\partial}{\partial t} \left [ \left(\delta (t)+\frac{j}{\pi t} \right)*u_k(t) \right] e^{-j \omega_k t}\right\rVert ^2 _2\\
&\text{subject to} && y(t) = \sum_{k=1}^{K} u_k(t)\\
\end{aligned} \label{eq:vmd}
\end{equation}
Here, $\left(\delta (t)+\frac{j}{\pi t} \right)*u_k(t)$ represents the analytic signal of $u_k(t)$. The $\delta (\cdot)$ and $*$ denote the Dirac delta function and the convolution, respectively.
The objective function in Equation (\ref{eq:vmd}) takes smaller values when the bandwidth around the center angular frequency, $\omega_k$, of each mode, $u_k(t)$, in the frequency distribution is small.
Please refer to Ref. \cite{Dragomiretskiy2014a} for further mathematical details for the estimation of the bandwidth in Equation (\ref{eq:vmd}).

In the actual algorithm, the following augmented Lagrangian is defined to loosen the constrained variational optimization problem (\ref{eq:vmd}).
That is, the solution to Equation (\ref{eq:vmd}) is sought as the saddle point of the following augmented Lagrangian (\ref{eq:vmd_lagran}):
\begin{equation}
{\cal L} \left( \{u_k\},\{\omega_k\},\lambda\right) =
\alpha \sum_{k=1}^{K} \left\lVert \frac{\partial}{\partial t} \left [ \left(\delta (t)+\frac{j}{\pi t} \right)*u_k(t) \right] e^{-j \omega_k t}\right\rVert ^2 _2 +
\left\lVert y(t) - \sum_{k=1}^{K} u_k(t) \right\rVert ^2 _2 +
\left< \lambda(t), y(t) - \sum_{k=1}^{K} u_k(t)\right> 
\label{eq:vmd_lagran}
\end{equation}
This problem (\ref{eq:vmd_lagran}) can be solved using the alternate direction method of multipliers (ADMM) \cite{Boyd2011}.
$K$ and $\alpha$ are parameters set by the user, allowing the adjustment of the number of modes and the penalty factor, respectively.

\subsection{Multivariate Variational Mode Decomposition}
MVMD is an extension of VMD that applies to multiple variables.
It expresses multivariate data as a sum of $K$ time-series patterns.
Here, the input data are denoted as $y_c(t)$, where $c=1,2,\cdots,C$, and subscript $c$ indicates the $c$-th variable.
The $k$-th MVMD mode corresponding to the $c$-th variable $y_c(t)$ is expressed as $u_{k,c}(t)$.
In MVMD, the MVMD mode, $u_{k,c}(t)$, and the center angular frequency, $\omega_k$, are computed as solutions to the following optimization problem:
\begin{equation}
\begin{aligned}
& \underset{\{u_{k,c}\},~\{\omega_k\}} {\text{minimize}} &&\sum_{c=1}^{C} \sum_{k=1}^{K} \left\lVert \frac{\partial}{\partial t} \left [ \left(\delta (t)+\frac{j}{\pi t} \right)*u_{k,c}(t) \right] e^{-j \omega_k t}\right\rVert ^2 _2\\
&\text{subject to} && y_c(t) = \sum_{k=1}^{K} u_{k,c}(t),~~~c=1,2,\cdots,C.\\
\end{aligned} \label{eq:mvmd}
\end{equation}
Thus, MVMD is a natural extension of VMD applicable to multivariate data; however, the formulation involves the center angular frequencies, $\omega_k$, taking common values regardless of $c$.
In other words, the $k$-th MVMD mode, $u_{k,c}(t)$, $c=1,2,\cdots,C$, extracts temporal fluctuations that vary with the same center angular frequency, $\omega_k$, from each variable, $y_c(t)$.

Similar to VMD, to determine $u_{k,c}(t)$ and $\omega_k$, we define the following augmented Lagrangian (\ref{eq:mvmd_lagran}) and find its saddle point using ADMM.
\begin{equation}
\begin{split}
{\cal L} \left( \{u_{k,c}\},\{\omega_k\},\lambda\right) &=
\alpha \sum_{c=1}^{C} \sum_{k=1}^{K} \left\lVert \frac{\partial}{\partial t} \left [ \left(\delta (t)+\frac{j}{\pi t} \right)*u_{k,c}(t) \right] e^{-j \omega_k t}\right\rVert ^2 _2 \\
&+ \sum_{c=1}^{C} \left\lVert y_c(t) - \sum_{k=1}^{K} u_{k,c}(t) \right\rVert ^2 _2 +
\sum_{c=1}^{C} \left< \lambda_c(t), y_c(t) - \sum_{k=1}^{K} u_{k,c}(t)\right> 
\label{eq:mvmd_lagran}
\end{split}
\end{equation}

\subsection{Alternate Direction Method of Multipliers for Multivariate Variational Mode Decomposition} 
\begin{algorithm}[hbt!]
\caption{Multivariate Variational Mode Decomposition}\label{alg:MVMD}
\begin{algorithmic}
    \Initialize{
        $\{ \hat{u}^1_{k,c} \},~\{ \hat{\omega}^1_{k} \},~\hat{\lambda}^1_c,~n \leftarrow 0,~\varepsilon \leftarrow 10^{-6}$
    }
    \Repeat
        \State $n \leftarrow n+1$
        \For{$k=1:K$}
            \For{$c=1:C$}
                \State \textit{Update mode} $\hat{u}_{k,c}$ \textit{for all} $\omega \ge 0$:
                \State \begin{equation*}
                    \hat{u}_{k,c}^{n+1}(\omega)  \leftarrow \frac{\hat{y}_c(\omega) - \sum_{i \neq k}\hat{u}^{n}_{i,c}(\omega) + \frac{\hat{\lambda}^n_c(\omega)}{2}}{1 + 2 \alpha (\omega -\omega^n_k)^2}
                \end{equation*}
            \EndFor
        \EndFor
        \For{$k=1:K$}
            \State \textit{Update center angular frequency} $\hat{\omega}_{k}$:\begin{equation*}
                \omega^{n+1}_k \leftarrow \frac{\sum_{c} \int_{0}^{\infty} \omega \left| \hat{u}_{k,c}^{n+1}(\omega) \right|^2 d\omega}{\sum_{c} \int_{0}^{\infty} \left| \hat{u}_{k,c}^{n+1}(\omega) \right|^2 d\omega}
            \end{equation*}
        \EndFor
        \For{$c=1:C$}
            \State \textit{Update} $\hat{\lambda}_{c}$ \textit{for all} $\omega \ge 0$:\begin{equation*}
                    \hat{\lambda}^{n+1}_c(\omega) \leftarrow \hat{\lambda}^{n}_c(\omega) + \tau \left(\hat{y}_c(\omega) - \sum_k \hat{u}^{n+1}_{k,c}(\omega) \right)
                \end{equation*}
        \EndFor
    
    \Until
        \textit{Convergence for all} $\omega \ge 0$: 
        \begin{equation*}
             \frac{\sum_k \sum_c \left|\hat{u}_{k,c}^{n+1}(\omega) - \hat{u}_{k,c}^{n}(\omega) \right|^2}{\sum_k \sum_c \left| \hat{u}_{k,c}^{n}(\omega) \right|^2} < \varepsilon
        \end{equation*}
    
\end{algorithmic}
\end{algorithm}

In this study, the saddle point of (\ref{eq:mvmd_lagran}) was solved using the ADMM method, referring to Ref. \cite{Rehman2019}.
ADMM involves solving sub-optimization problems and iteratively updating the modes, center angular frequencies, and Lagrange multipliers.
The solutions to the sub-optimization problems are given below.
\begin{equation}
    \hat{u}_{k,c}^{n+1}(\omega)  \leftarrow \frac{\hat{y}_c(\omega) - \sum_{i \neq k}\hat{u}^{n}_{i,c}(\omega) + \frac{\hat{\lambda}^n_c(\omega)}{2}}{1 + 2 \alpha (\omega -\omega^n_k)^2}
\end{equation}
\begin{equation}
    \omega^{n+1}_k \leftarrow \frac{\sum_{c} \int_{0}^{\infty} \omega \left| \hat{u}_{k,c}^{n+1}(\omega) \right|^2 d\omega}{\sum_{c} \int_{0}^{\infty} \left| \hat{u}_{k,c}^{n+1}(\omega) \right|^2 d\omega}
\end{equation}
\begin{equation}
    \hat{\lambda}^{n+1}_c(\omega) \leftarrow \hat{\lambda}^{n}_c(\omega) + \tau \left(\hat{y}_c(\omega) - \sum_k \hat{u}^{n+1}_{k,c}(\omega) \right)
\end{equation}
where $\hat{\cdot}$ represents the frequency-domain expression of the corresponding variable, $n$ denotes the iteration number, and $\tau$ is the parameter for controlling the step size to update the Lagrangian multiplier.
The algorithm is summarized in Algorithm 1.
In this study, besides the convergence criterion in Algorithm 1, we confirmed that the reconstruction error, computed by the following equation (\ref{eq:reconst_error}) was less than $10^{-6}$:
\begin{equation}
    E = \frac{\sum_{c} \left\lVert y_c(t) - \sum_{k} u_{k,c}(t) \right\rVert_2^2}{\sum_{c} \left\lVert y_c(t) \right\rVert_2^2}\label{eq:reconst_error}
\end{equation} 

\subsection{Variational Mode Decomposition--Based Nonstationary Coherent Structure Analysis}

\subsubsection{Input data}
The data under analysis are high-dimensional time-series data, such as numerical fluid simulation data or experimental images.
We analyze the fluctuating component obtained by subtracting the time-averaged field from the instantaneous values:
\begin{equation}
    q'(\bm x, t) = q(\bm x, t) - \bar{q}(\bm x) 
\end{equation}
Here, $q'(\bm x, t)$, $q(\bm x, t)$, and $\bar{q}(\bm x)$ represent fluctuating, instantaneous, and time-average components, respectively.
The fluctuating component of the flow field at time $t=t_j$ is denoted as a column vector, $q'(t_j)$.
In fluid analysis, the flow field, $q'(t_j)$, at each time instant is also referred to as a snapshot.
The input data for the VMD-NCS analysis can be represented as a matrix composed by arranging these snapshots, $q'(t_j)~(j=1,\cdots,m)$:
\begin{equation}
    Q' =
\begin{bmatrix}
   | & | & & | \\
   q'(t_1)  & q'(t_2) &\cdots & q'(t_m) \\
   | & | & & | 
\end{bmatrix}
\end{equation}
In this study, the number of rows is denoted as $d$, and the number of columns (i.e., the number of snapshots) is denoted as $m$.
Thus, $Q'$ is a $d \times m$ matrix.

\subsubsection{Dimensionality reduction}
We perform dimensionality reduction of the input data using POD.
Let $P$ be the POD basis of $Q'$:
\begin{equation}
    P =
\begin{bmatrix}
   | & | & & | \\
   \phi_1  & \phi_2 &\cdots & \phi_r \\
   | & | & & | 
\end{bmatrix}
\end{equation}
The low-dimensional representation of $Q'$ is computed as follows:
\begin{equation}
    \tilde{Q} = P^{\mathsf{T}} Q' =
\begin{bmatrix}
   \tilde{q}_{1}(t_1) & \cdots  & \tilde{q}_{1}(t_m)\\
   \vdots  & \ddots & \vdots\\
   \tilde{q}_{r}(t_1) & \cdots  & \tilde{q}_{r}(t_m)
\end{bmatrix}
\end{equation}
Here, $r$ ($r \ll d$) is the truncation rank in the POD method.
$\tilde{Q}$ represents $Q'$ in a low-dimensional form as the expansion coefficients of POD.
Thus, snapshot $q'(\bm x, t)$ at time $t$ is represented using the POD basis by the following equation:
\begin{equation}
    q'(\bm x, t) =  \sum_{i=1}^{r} \tilde q_i(t) \phi_i(\bm x) + \epsilon_{POD}(\bm x, t) \label{eq:lowdim_q}
\end{equation}
In the above equation (\ref{eq:lowdim_q}), in correspondence with the conventional form of modal decomposition (\ref{eq:mode_decomp}) and for clarity, position $\bm x$ and time $t$ are explicitly indicated.

\subsubsection{Intrinsic coherent structure}
In the VMD-NCS analysis, MVMD is applied to $\tilde{Q}$ to obtain MVMD mode $u_{k,i}(t)$ and center angular frequency $\omega_k$, ($k=1,\cdots,K$).
For example, $\tilde q_i(t)$ is decomposed into $u_{k,i}(t)$, as follows:
\begin{eqnarray}
    \tilde q_i(t) =  \sum_{k=1}^{K} u_{k,i}(t) \label{eq:MVMDofPOD}
\end{eqnarray}
That is, by applying MVMD, this method decomposes the POD expansion coefficients into separate temporal patterns, each associated with a center frequency corresponding to $k$.
Then, the spatiotemporal patterns represented by the $k$-th MVMD mode and the POD basis, namely the nonstationary (or stationary) coherent structure extracted using the VMD-NCS analysis, are computed by the following equation:
\begin{eqnarray}
    \psi_k(\bm x, t) = \sum_{i=1}^{r} u_{k,i}(t) \phi_i(\bm x) \label{eq:ics}
\end{eqnarray}
In this study, $\psi_k(\bm x, t)$ is referred to as the intrinsic coherent structure (ICS).
Snapshot $q'(\bm x, t)$ at time $t$ can be represented using ICS, as follows:
\begin{equation}
    q'(\bm x, t) =  \sum_{k=1}^{K}  \psi_k(\bm x, t) + \epsilon_{ICS}(\bm x, t) 
    \label{eq:sum_ics}
\end{equation} 
Note that, Equations (\ref{eq:ics})--(\ref{eq:sum_ics}) explicitly state that the ICS ($\psi_k(\bm x, t)$) is a function of position $\bm x$ and time $t$.
Thus, in contrast to the case in the conventional modal decomposition (\ref{eq:mode_decomp}), the spatial distribution of the ICS can vary with time.

\subsubsection{Amplitude of the intrinsic coherent structure}
The amplitude of the $k$-th ICS at time $t = t_j$, $\| \psi_k(t_j) \|$, can be easily computed due to the orthogonality of the POD basis, $P$, as follows:
\begin{equation}
    \| \psi_k(t_j) \| = \sqrt{\sum_{i=1}^{r} u_{k,i}(t_j)^2}
\end{equation}
Additionally, we define the time-averaged amplitude, $\overline{\| \psi_k \|}$, by the following equation:
\begin{equation}
    \overline{\| \psi_k \|} = \frac{1}{m} \sum_{j=1}^{m} \| \psi_k (t_j) \|
\end{equation}

\subsubsection{Setting parameters}
In MVMD, there are two user-setting parameters: $\alpha$, which regulates the penalty term in Equation (\ref{eq:mvmd_lagran}), and $K$, which represents the number of ICSs.
A large $\alpha$ results in a narrow bandwidth effect. Accordingly, the extracted ICSs capture phenomena with relatively narrow frequencies, i.e., high periodicity, while failing to represent nonperiodic variations.
Conversely, reducing parameter $\alpha$ weakens the periodicity.
If $\alpha$ becomes overly small, the temporal coherence of ICSs may be lost.
Notably, a high $K$ and a low $\alpha$ increase the difficulty of comprehending the phenomena.
Conversely, if the $K$ is overly low or $\alpha$ is overly high, it might be unfeasible to represent the input data through the superposition of ICSs, increasing the reconstruction error (\ref{eq:reconst_error}).
Therefore, it is important to set parameters $K$ and $\alpha$ within a range that ensures the extracted ICSs are easily interpretable while maintaining a sufficiently small reconstruction error.
The parameter dependency when applying the VMD-NCS analysis to actual flow fields is discussed in Sections \ref{sec:TWBC} and \ref{sec:FAPA}.

\section{Transient wake behind a cylinder}\label{sec:TWBC}
\subsection{Problem Setting and Input data}
As the first validation case, the transient growth phenomena around a two-dimensional square cylinder were analyzed.
The input data consisted of temporal distributions of $x$- and $y$-direction velocities obtained through a two-dimensional numerical fluid simulation.
In the numerical fluid simulation, we first employed the selective frequency damping method \cite{Akervik2006} to compute an unstable steady-state solution, which exhibited twin vortices behind the cylinder.
Subsequently, by evolving this steady-state solution from time $t=0$ as the initial condition, temporal data capturing the transient growth phenomena from the unstable steady-state to the periodic shedding of Kármán vortices were acquired.
The freestream Reynolds number based on the length of one side of the square cylinder was 120, and the freestream Mach number was 0.2.
For details on this numerical fluid simulation, please refer to Ref. \cite{Ohmichi2021}.

\begin{figure}[hbt!]
    \centering
    \includegraphics[width=.5\textwidth]{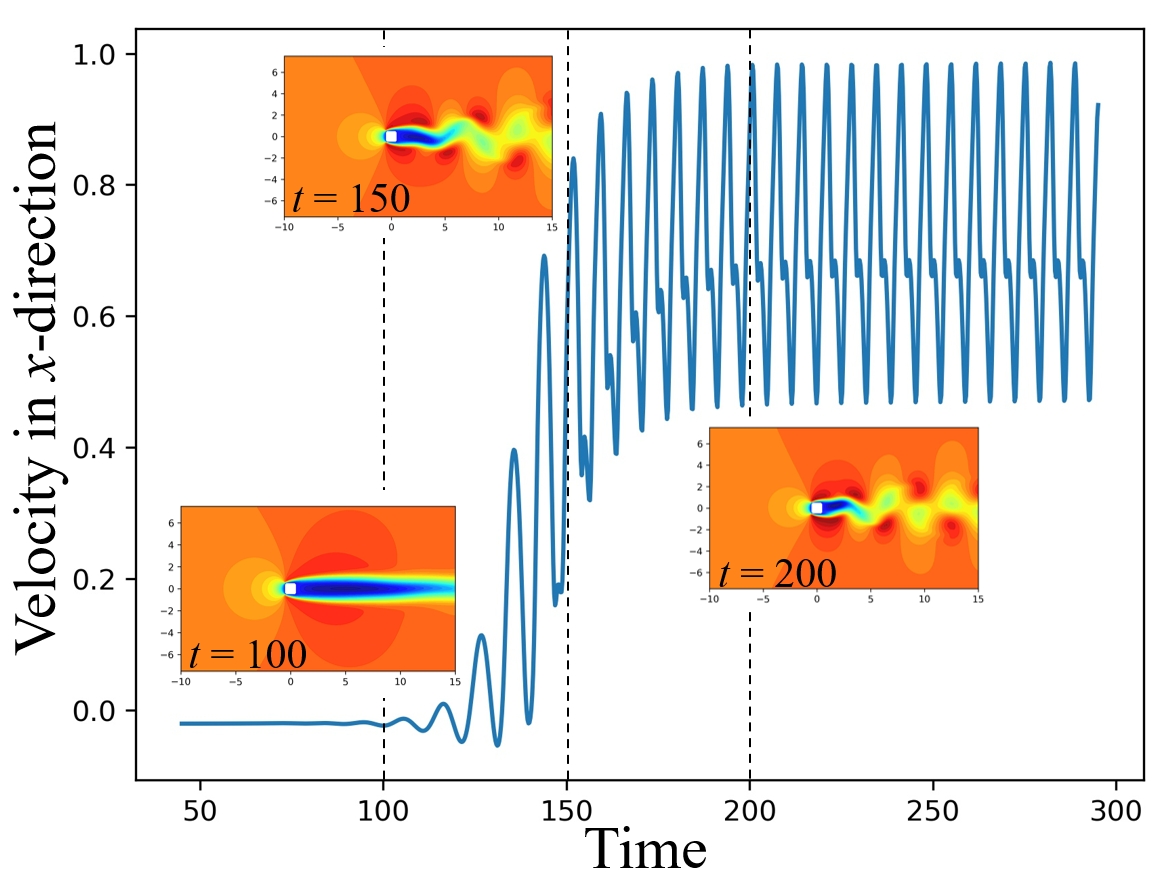}
    \caption{Time history of the velocity in the $x$-direction at $(x,~y)=(5,~0.5)$ and spatial distributions of the velocity in the $x$-direction at time $t = 100$, 150, and 200.}
    \label{fig:cylinder_vel_field}
\end{figure}

To illustrate an overview of the flow field of this analysis, Fig. \ref{fig:cylinder_vel_field} shows the temporal evolution of the $x$-direction velocity at position $(x,~y)=(5,~0.5)$ in the wake region of a square cylinder and the spatial distribution of the $x$-direction velocity at multiple times during the transient growth occurring at $100 \leq t \leq 200$.
The positions, times, and velocities are non-dimensionalized based on the length of a side of the cylinder and the freestream velocity.
A slight variation is observed until around $t=100$, beyond which the fluctuation gradually and evidently goes through a transient growth phase, leading to periodic variations corresponding to the shedding of the periodic Kármán vortices.

As input datasets for the VMD-NCS analysis, the mean-subtracted velocity field around the cylinder within the ranges of $x = [-10,~15]$ and $y = [-7.5,~7.5]$ was sampled at equally spaced grid points of $251 \times 151$.
The number of snapshots was 1000, and the time interval between snapshots was $\Delta t = 0.25$.
These input datasets correspond to the flow field between $t=45$ and $t=295$.
Therefore, these datasets encompass both the transient growth between $100 \leq t \leq 200$ and the periodic variations after $t>200$.

The truncation rank via POD was set to $r=23$ to ensure that the sum of contributions (variances) exceeds 99.9\%.
Although not shown in this paper, it was confirmed that the result remains consistent even with $r=10$.

\subsection{Results and Discussion}
\subsubsection{Extracted coherent structures ($K = 5,~\alpha = 100$)}
\begin{figure}[hbt!]
    \centering
    \includegraphics[width=.85\textwidth]{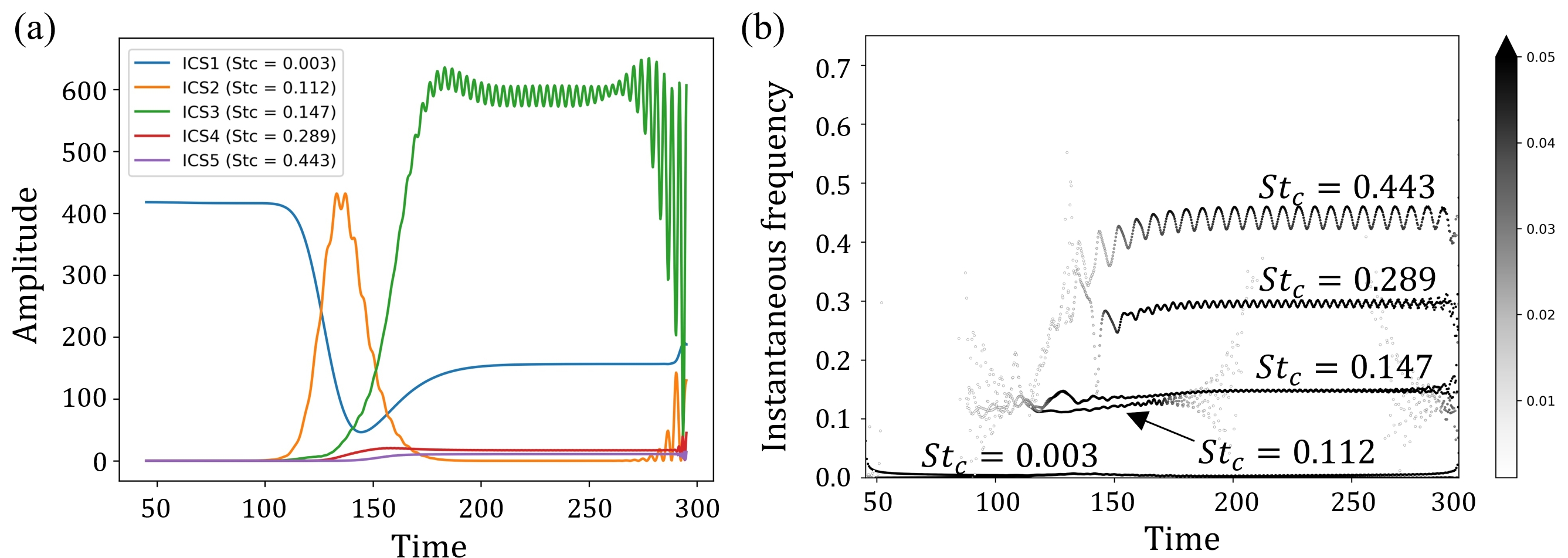}
    \caption{(a) Time history of the ICS amplitudes and (b) time-frequency distribution of the $x$-direction velocity component at position $(x,~y)=(5,~0.5)$ for each ICS. The shade of the points in (b) represents the amplitude. $K = 5,~\alpha = 100$.}
    \label{fig:cylinder_amp_instfreq_alpha100}
\end{figure}

\begin{figure}[hbt!]
    \centering
    \includegraphics[height=15cm]{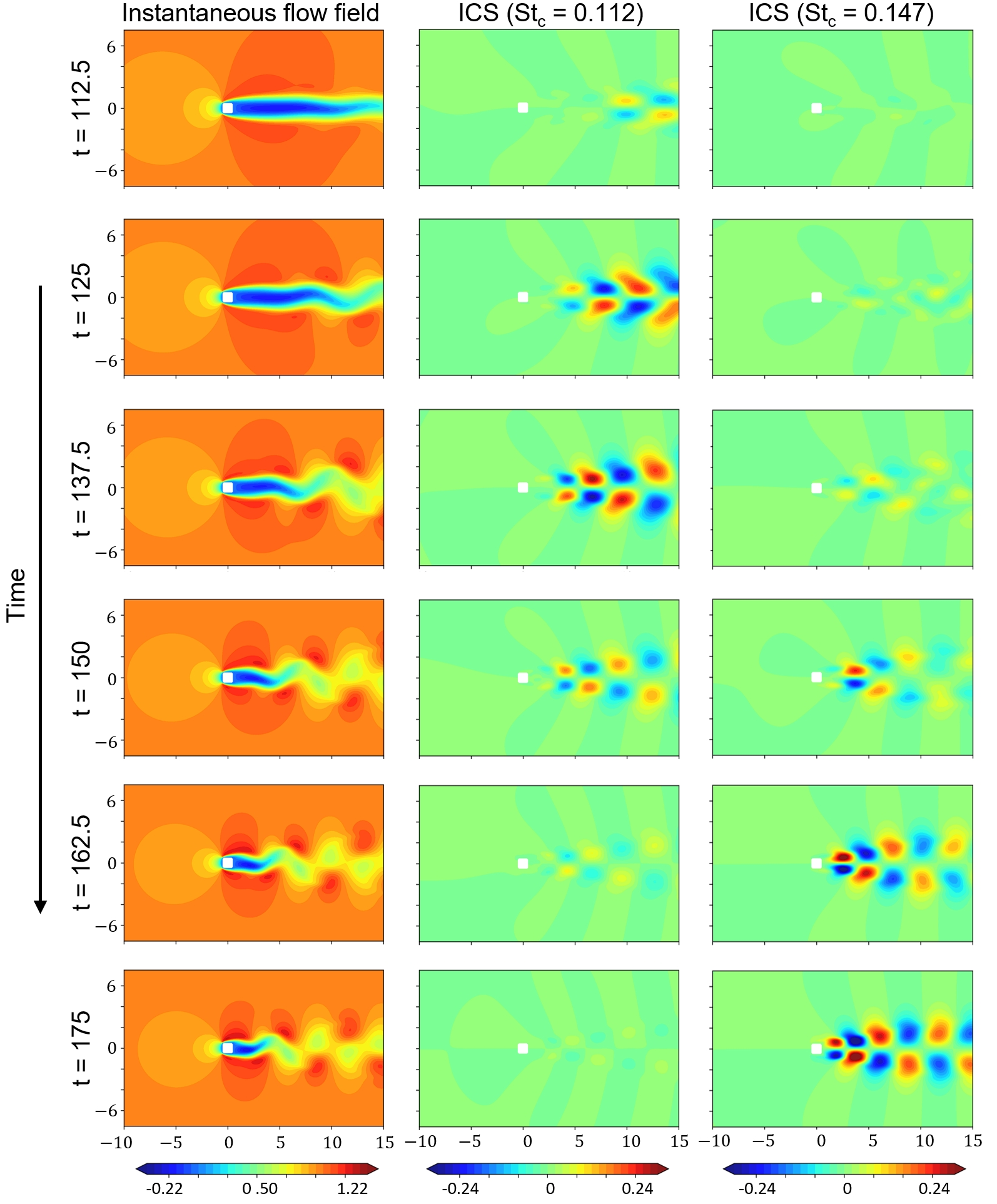}
    \caption{Instantaneous spatial distributions of the input flow field (with mean components) and the ICSs during the transient growth period. Input flow field (left), ICS of $St_c=0.112$ (middle), and ICS of $St_c=0.147$ (right). The $x$-direction velocity component is shown.}
    \label{fig:cylinder_ICS_shape1}
\end{figure}

\begin{figure}[hbt!]
    \centering
    \includegraphics[height=15cm]{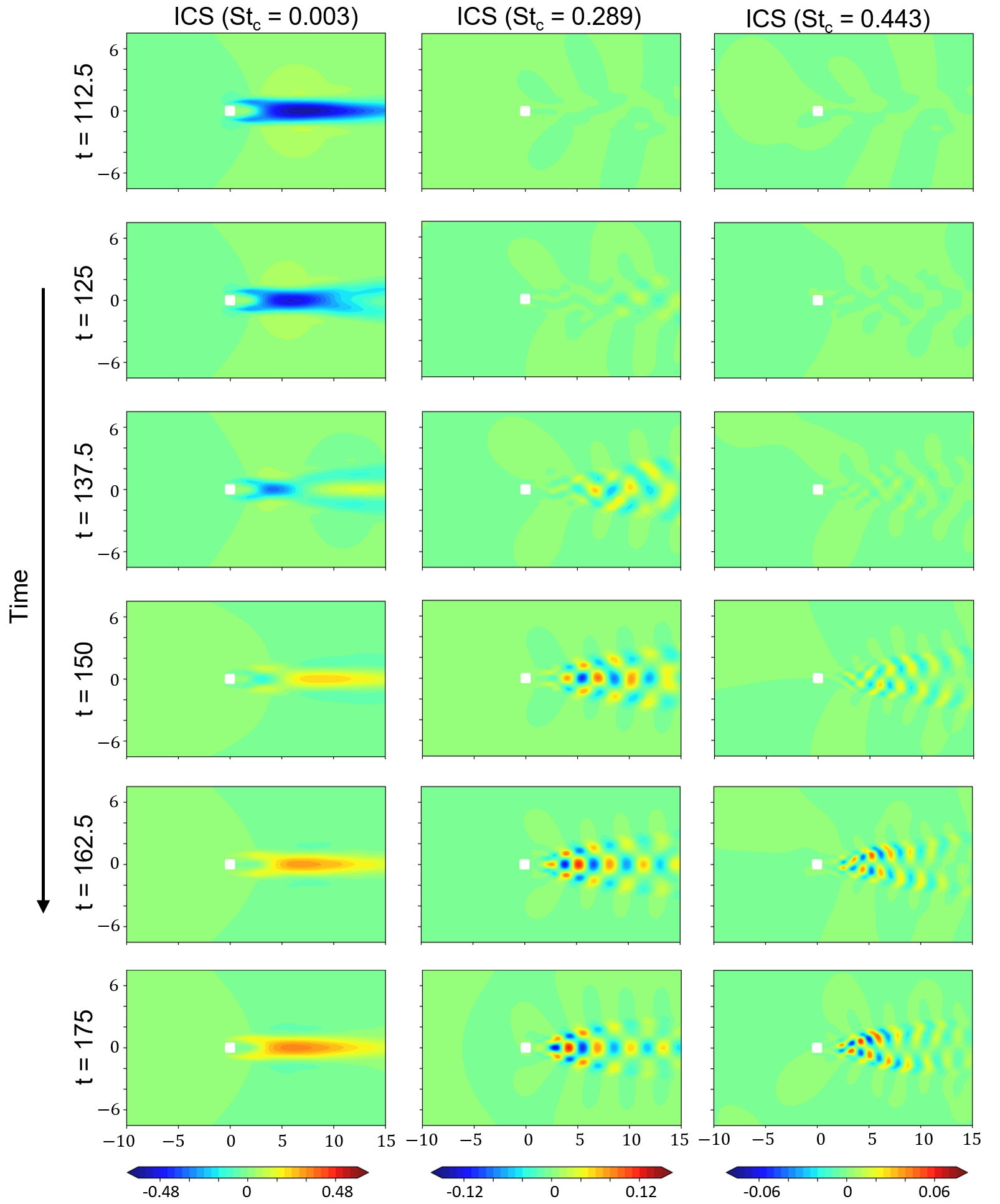}
    \caption{Instantaneous spatial distributions of the ICSs during the transient growth period. ICS of  $St_c=0.003$ (left), $St_c=0.289$ (middle), and $St_c=0.443$ (right). The $x$-direction velocity component is shown.}
    \label{fig:cylinder_ICS_shape2}
\end{figure}

First, the ICSs at $K = 5$ and $\alpha = 100$ are evaluated. Fig. \ref{fig:cylinder_amp_instfreq_alpha100} shows the temporal evolution of the ICS amplitudes, $\|\bm \psi_k(t)\|$, and the time-frequency distribution obtained by the Hilbert transform of the ICSs with respect to the $x$-direction velocity at position $(x,~y)=(5,~0.5)$.
Setting $K=5$ yields five ICSs.
Among these obtained ICSs, three exhibited significant amplitudes at $St_c = 0.003$, 0.112, and 0.147 (Fig. \ref{fig:cylinder_amp_instfreq_alpha100}(a)).
Here, $St_c$ denotes the nondimensionalized center frequency of each ICS.
From Fig. \ref{fig:cylinder_amp_instfreq_alpha100}(a), it is evident that the ICS of $St_c = 0.112$ became dominant during the transient growth period of the flow (i.e., it had a large amplitude), whereas the ICS of $St_c = 0.147$ became dominant during the subsequent periodic shedding of the Kármán vortices.
The spatial distribution and amplitude variation of these ICSs with time are depicted in Fig. \ref{fig:cylinder_ICS_shape1}.
The figure illustrates the growth of the ICS of $St_c = 0.112$, representing transient growth, followed by the appearance of the ICS of $St_c = 0.147$, representing the periodic shedding of the Kármán vortices. The figure also shows the subsequent disappearance of the ICS of $St_c = 0.112$.
Moreover, for the ICS of $St_c = 0.112$, the vortex shedding initially manifested at the far-wake region of the cylinder ($t = 112.5$) and gradually approached the near-wake region of the cylinder.
Note that such spatial-distribution changes cannot be represented as a single mode using conventional methods, including POD and DMD, which are in the form of Equation (\ref{eq:mode_decomp}).
This result demonstrates that these spatial-distribution changes can be represented as a single ICS using the VMD-NCS analysis method, highlighting its advantage over conventional methods in capturing the time-varying nature of nonstationary phenomena.

Furthermore, on examining the results of the time-frequency analysis (Fig. \ref{fig:cylinder_amp_instfreq_alpha100}(b)), it was confirmed that each ICS exhibited time-varying frequency appearing around their center frequencies, and the frequency and amplitude gradually varied over time.

Fig. \ref{fig:cylinder_ICS_shape2} illustrates the spatial distributions of ICSs for $St_c = 0.003$, 0.289, and 0.443.
The ICS of $St_c=0.003$ represents the gradual temporal changes in the background flow.
As shown in Fig. \ref{fig:cylinder_amp_instfreq_alpha100}(a), as the amplitude of the ICS of $St_c = 0.003$ decreased, the ICS of $St_c = 0.112$ increased, and the early period of transient growth was predominantly expressed by the behaviors of these ICSs.
The ICSs of $St_c = 0.289$ and 0.443 correspond to two and three times the frequency of the Kármán vortices, respectively, capturing finer structures but with reduced amplitudes.

\subsubsection{Comparison of the proper orthogonal decomposition coefficients and multivariate variational mode decomposition modes ($K = 5,~\alpha = 100$)}

\begin{figure}[hbt!]
    \centering
    \includegraphics[width=.8\textwidth]{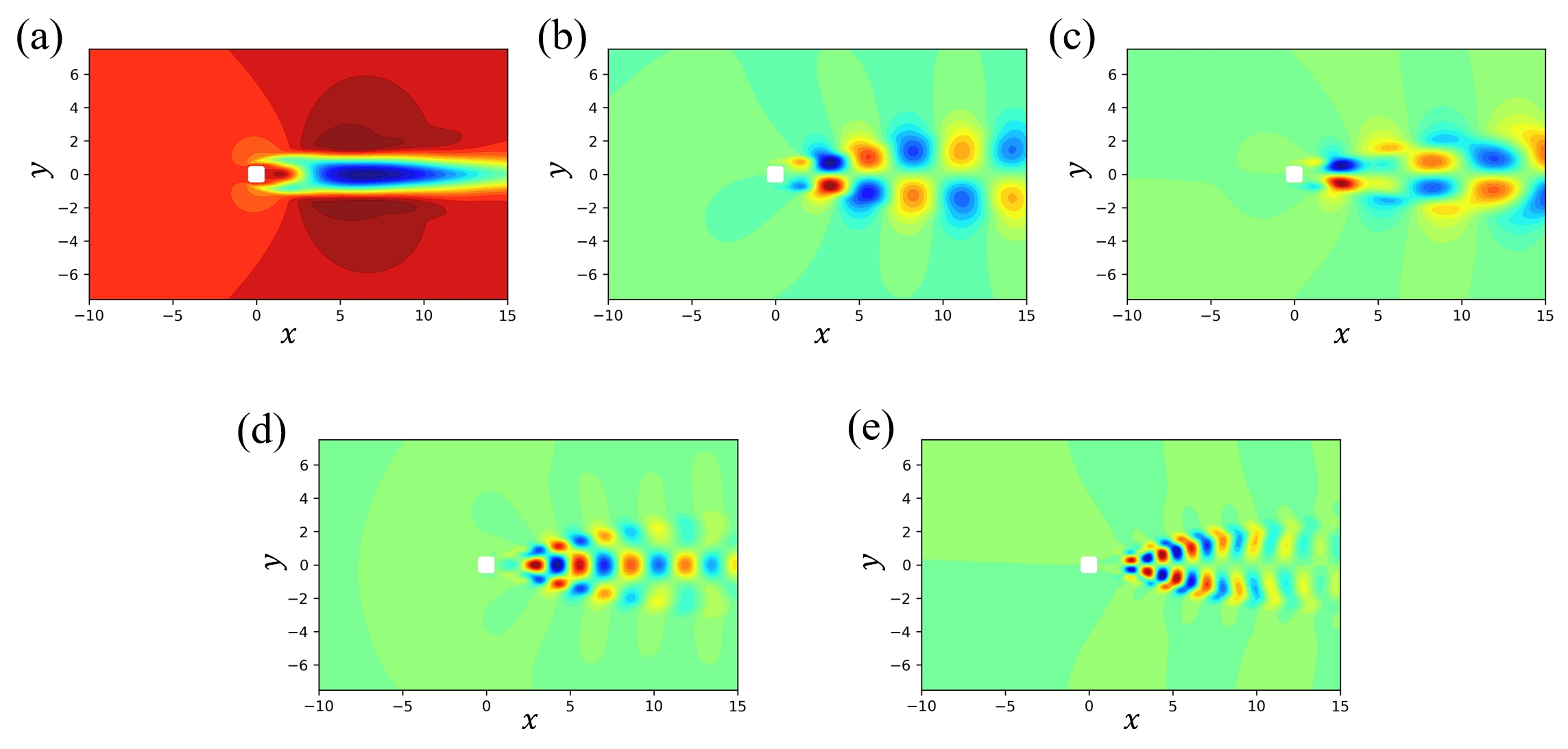}
    \caption{The spatial distribution of the $x$-direction velocity component for the (a) first, (b) third, (c) fifth, (d) seventh, and (e) ninth POD modes.}
    \label{fig:pod_mode_shapes}
\end{figure}

\begin{figure}[hbt!]
    \centering
    \includegraphics[width=.99\textwidth]{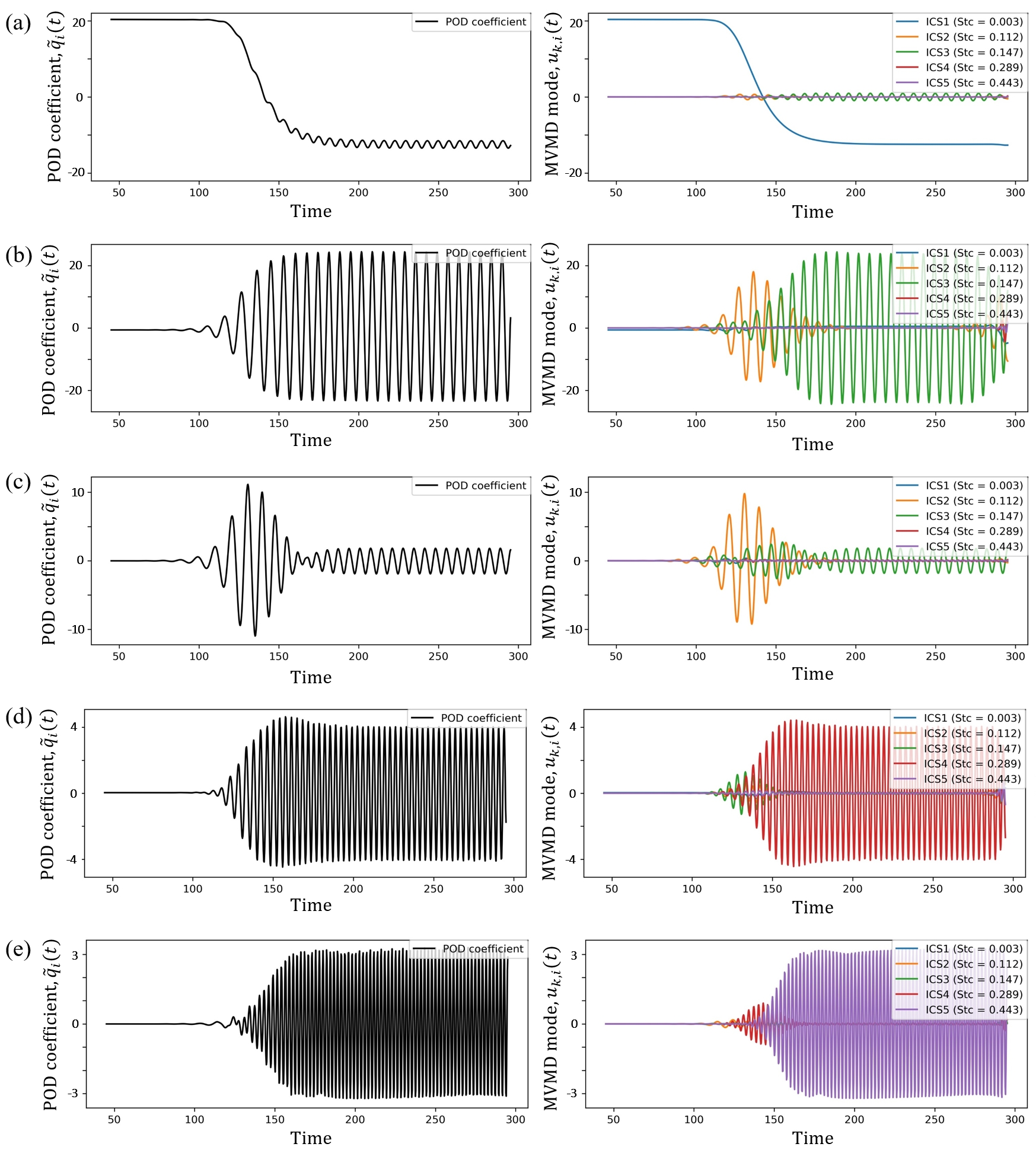}
    \caption{The temporal evolution of the expansion coefficients (left) and MVMD modes obtained by decomposing the expansion coefficients (right) for the (a) first, (b) third, (c) fifth, (d) seventh, and (e) ninth POD modes.}
    \label{fig:cylinder_POD_MVMD_coeff}
\end{figure}

Fig. \ref{fig:pod_mode_shapes} shows the spatial distributions of the POD modes $\bm \phi_i$.
Additionally, Fig. \ref{fig:cylinder_POD_MVMD_coeff} presents the temporal evolution of the POD coefficients $\tilde{q}_i(t)$, and the MVMD modes $u_{k,i}(t)$ obtained by decomposing these POD coefficients (\ref{eq:MVMDofPOD}).
The spatial distribution of the POD modes appears to resemble the distribution of ICSs at some time instances shown in Figs. \ref{fig:cylinder_ICS_shape1} and \ref{fig:cylinder_ICS_shape2}.
However, on observing the temporal evolution of the POD expansion coefficients, it is evident that each POD mode comprises components of various phenomena, including transient changes, periodic variations, and gradual temporal changes in the background flow.
For instance, the coefficients of the first POD mode contained a mixture of slowly varying components and periodic fluctuations.
Similarly, coefficients of other modes exhibited variations representing the transient growth of the flow field and periodic fluctuations within a single mode.
Conversely, the MVMD modes effectively decomposed these into temporal variations, corresponding to individual phenomena.
In the VMD-NCS analysis, the MVMD method enables the extraction of individual phenomena from the temporal evolution of the POD mode expansion coefficients, and superimposing the POD modes for each extracted phenomena enables the extraction of spatiotemporal patterns of nonstationary phenomena as ICSs.

\subsubsection{Parameter dependency}
\begin{figure}[hbt!]
    \centering
    \includegraphics[width=.99\textwidth]{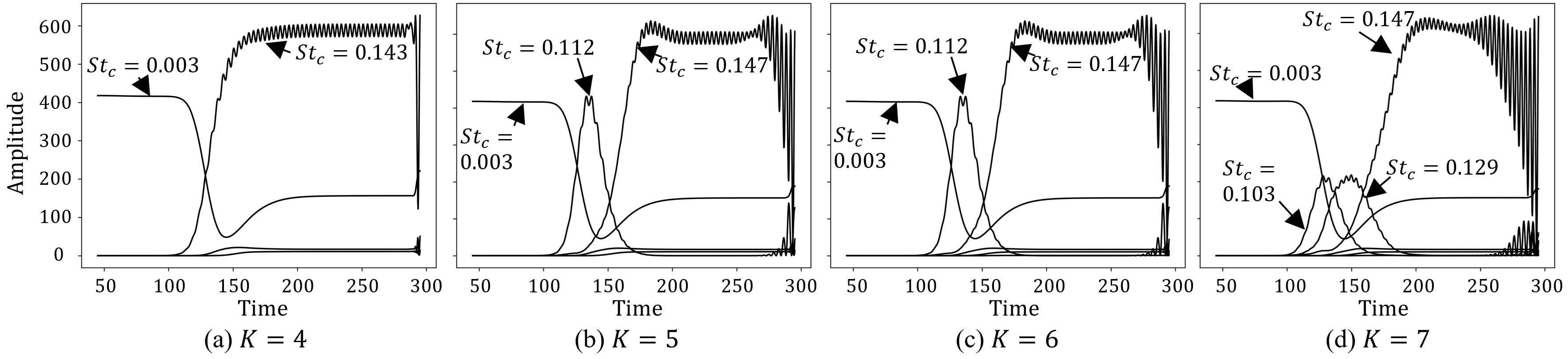}
    \caption{Influence of parameter $K$; time history of the ICS amplitudes; $\alpha=1000$.}
    \label{fig:cylinder_K_effect_amplitude}
\end{figure}

\begin{figure}[hbt!]
    \centering
    \includegraphics[width=.99\textwidth]{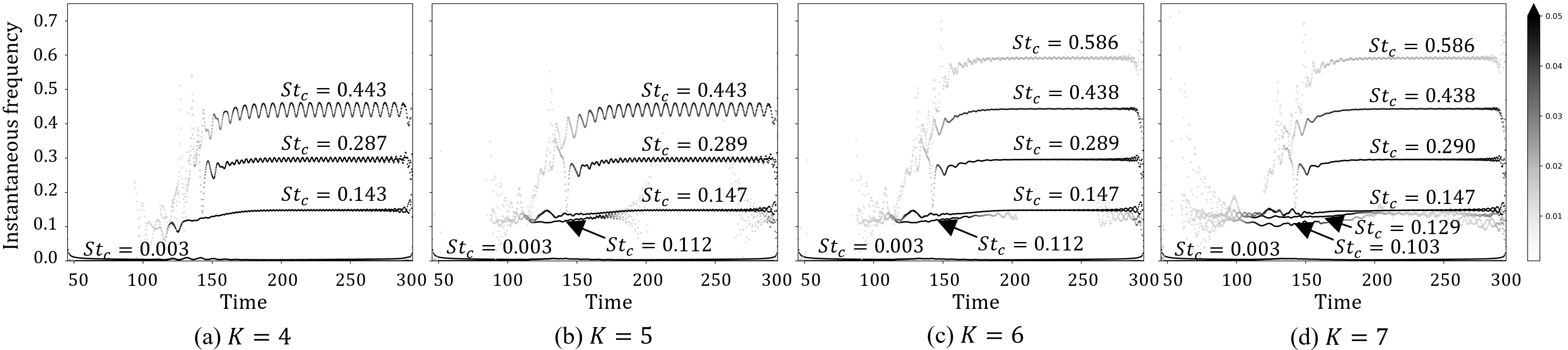}
    \caption{Influence of parameter $K$; time-frequency distribution of the $x$-direction velocity component at position $(x,~y)=(5,~0.5)$ for each ICS. The shade of the points represents the amplitude. $\alpha=1000$.}
    \label{fig:cylinder_K_effect_instfreq}
\end{figure}

The time evolution of the ICS amplitudes and the time-frequency distribution of the $x$-direction velocity component at position $(x,~y)=(5,~0.5)$ for each ICS, with varying the number of ICSs ($K=4$, 5, 6, and 7), are depicted in Figs. \ref{fig:cylinder_K_effect_amplitude} and \ref{fig:cylinder_K_effect_instfreq}, respectively.
Parameter $\alpha$ was fixed at 1000.
For $K=4$, the ICS corresponding to the Kármán vortex ($St_c=0.147$) and the ICS of $St_c=0.112$, representing the transient fluctuation observed for $K=5$, were not distinct; instead, they merged.
For $K=6$, a new ICS corresponding to four times the frequency of the Kármán vortex ($St_c=0.586$) emerged.
For $K=7$, in addition to the aforementioned ICSs, the ICS of $St_c=0.112$ split into two ICSs.
As shown in Fig. \ref{fig:cylinder_K_effect_instfreq}, the variations in the instantaneous frequency for the ICSs with $St_c=0.289$ and $St_c=0.438$ were noticeable at $K=4$ and 5, but were suppressed at $K=6$, exhibiting a behavior akin to the steady frequency behavior.

\begin{figure}[hbt!]
    \centering
    \includegraphics[width=.85\textwidth]{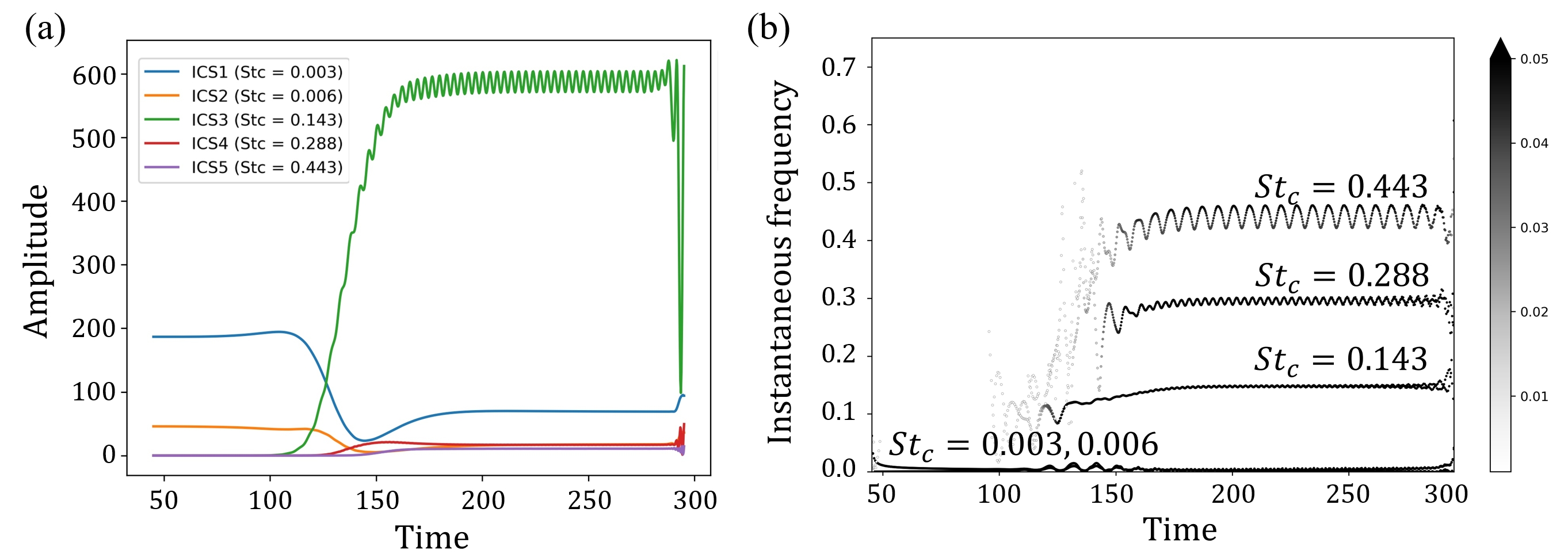}
    \caption{(a) Time history of the ICS amplitudes and the (b) time-frequency distribution of the $x$-direction velocity component at position $(x,~y)=(5,~0.5)$ for each ICS. The shade of the points in (b) represents the amplitude. $K = 5,~\alpha = 1$.}
    \label{fig:cylinder_amp_instfreq_alpha001}
\end{figure}

Next, we investigated the dependency on parameter $\alpha$, which regulates the penalty term related to the frequency bandwidth of the ICS.
Fig. \ref{fig:cylinder_amp_instfreq_alpha001} shows the amplitude and time-frequency distribution of the ICSs for $K=5$ and $\alpha=1$.
When compared to the results for $K=5$ and $\alpha=100$ (Fig. \ref{fig:cylinder_K_effect_amplitude}),
setting $\alpha=1$ which represents a weaker penalty, leads to the disappearance of the ICS with $St_c=0.112$ which  captured the transient fluctuations.
Instead, a new ICS with $St_c=0.006$ emerged.
No significant changes in the obtained ICSs were observed for $\alpha=10$, 100, and 1000 (i.e., same as the results shown in Fig. \ref{fig:cylinder_K_effect_amplitude}).
Thus, in this analysis, the dependency on $\alpha$ was not particularly pronounced.
Note that, for $\alpha>1000$, it was not possible to obtain solutions with a sufficiently small reconstruction error (\ref{eq:reconst_error}).

In this analysis, parameter $K$ exerted a greater impact on the obtained ICSs than $\alpha$.
However, when $\alpha$ is overly small and the penalty imposed on each ICS is weak, the temporal coherence of the ICS may be lost, complicating the extraction of physically meaningful phenomena.
While the investigation of the parameter dependency in the VMD-NCS analysis remains a future research topic, adjusting the parameters by fixing $\alpha$ to a large value within a range where the reconstruction error (\ref{eq:reconst_error}) remains sufficiently small and varying $K$ could be considered a good approach.

\section{Flow around a pitching airfoil}\label{sec:FAPA}
\subsection{Problem Setting and Input Data}
In this section, the analysis of the low-Reynolds-number flow around a pitching airfoil is reported.
The analyzed data were obtained from a publicly accessible database \cite{Towne2023}.
The data comprised a two-dimensional flow field around a flat-plate airfoil with periodic pitching motion.
The freestream Reynolds number based on the airfoil chord length was 100.
The angle of attack of the airfoil oscillates according to the equation:
\begin{equation}
    a(t) = a_0 - a_P {\rm sin}(2 \pi f_P t). \label{eq:phase},
\end{equation}
where $a_0 = 30.0$ deg and $a_P = 5.0$ deg, indicating that the angle of attack varies sinusoidally within a range of $\pm 5$ deg centered around 30 deg.
Here, $t$ represents the dimensionless time, and $f_P$ is the dimensionless frequency, set at $f_P=0.05$ for this analysis.

According to Ref. \cite{Towne2023}, vortex shedding occurs at a frequency of $St=0.2418$ even without pitching at an angle of attack of 30 deg.
This flow field exhibits primarily two fluctuating phenomena: vortex shedding and fluctuation due to pitching, resulting in a quasi-periodic flow field.
This analysis aims to investigate the mechanism through which these phenomena are captured by VMD-NCS analysis.

The mean-subtracted vorticity field around a flat-plate airfoil was employed as the input datasets for the VMD-NCS analysis.
Each snapshot had equally spaced $599 \times 299$ grid points,
and the number of snapshots was 1001.
The time interval between the snapshots was $\Delta t = 0.1$.
For further details of the datasets, please refer to Ref. \cite{Towne2023}.

The truncation rank for the POD was set to $r=54$, ensuring that the sum of contributions exceeds 99.9\%.
While not shown here, it has been confirmed that even at $r=25$, the results remained unchanged.

\subsection{Results and Discussion}
\subsubsection{Intrinsic coherent structure frequency distribution}
\begin{figure}[hbt!]
    \centering
    \includegraphics[width=.85\textwidth]{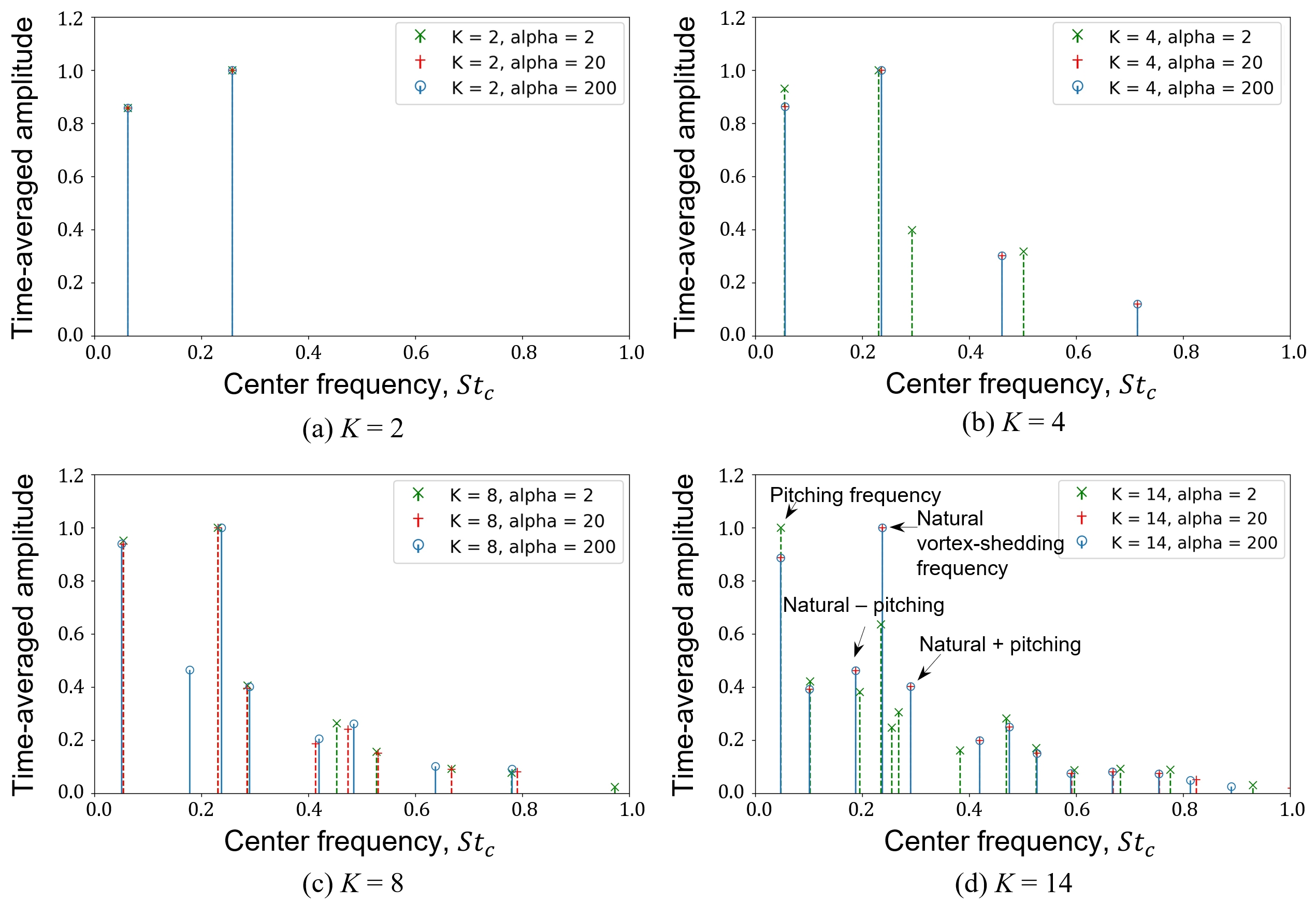}
    \caption{Influence of parameters $K$ and $\alpha$ on the center frequency and mean amplitude of the ICSs.}
    \label{fig:pitching_ics_amp}
\end{figure}
The center frequencies and time-averaged amplitudes of the ICS with varying parameters $K$ and $\alpha$ are shown in Fig. \ref{fig:pitching_ics_amp}.
The amplitudes are normalized so that the largest amplitude is unity in the figure.
For $K=2$ (Fig. \ref{fig:pitching_ics_amp}(a)), a similar distribution is observed for all $\alpha$ cases.
The center frequencies of the two ICSs were $St_c = 0.063$ and $St_c = 0.258$.
These frequencies were close to the pitching frequency ($f_p = 0.05$) and the vortex shedding frequency ($St = 0.2418$), respectively, indicating the successful extraction of phenomena related to their respective variations.
As the number of the ICSs increased, the dependence on $\alpha$ appeared.
For $K=4$, there was little difference between the results for $\alpha = 20$ and 200, where, besides the ICS corresponding to the pitching and vortex shedding frequencies, the second and third harmonics of the vortex shedding frequency ($St_c = 0.461$ and $St_c = 0.714$, respectively) were extracted.
For the smallest value (corresponding to the weakest penalty), $\alpha = 2$, a different ICS emerged around the vortex shedding frequency, resulting in a different frequency distribution.
Upon further increasing the $K$, the ICS associated with the main vortex shedding frequency and its second harmonic were further divided, generating ICSs at frequencies plus and minus the pitching frequency, as shown in Fig. \ref{fig:pitching_ics_amp}(d).
At $\alpha = 2000$, the reconstruction error (\ref{eq:reconst_error}) increased for all $K$ values, and a convergent solution could not be obtained.

\begin{figure}[hbt!]
    \centering
    \includegraphics[width=.45\textwidth]{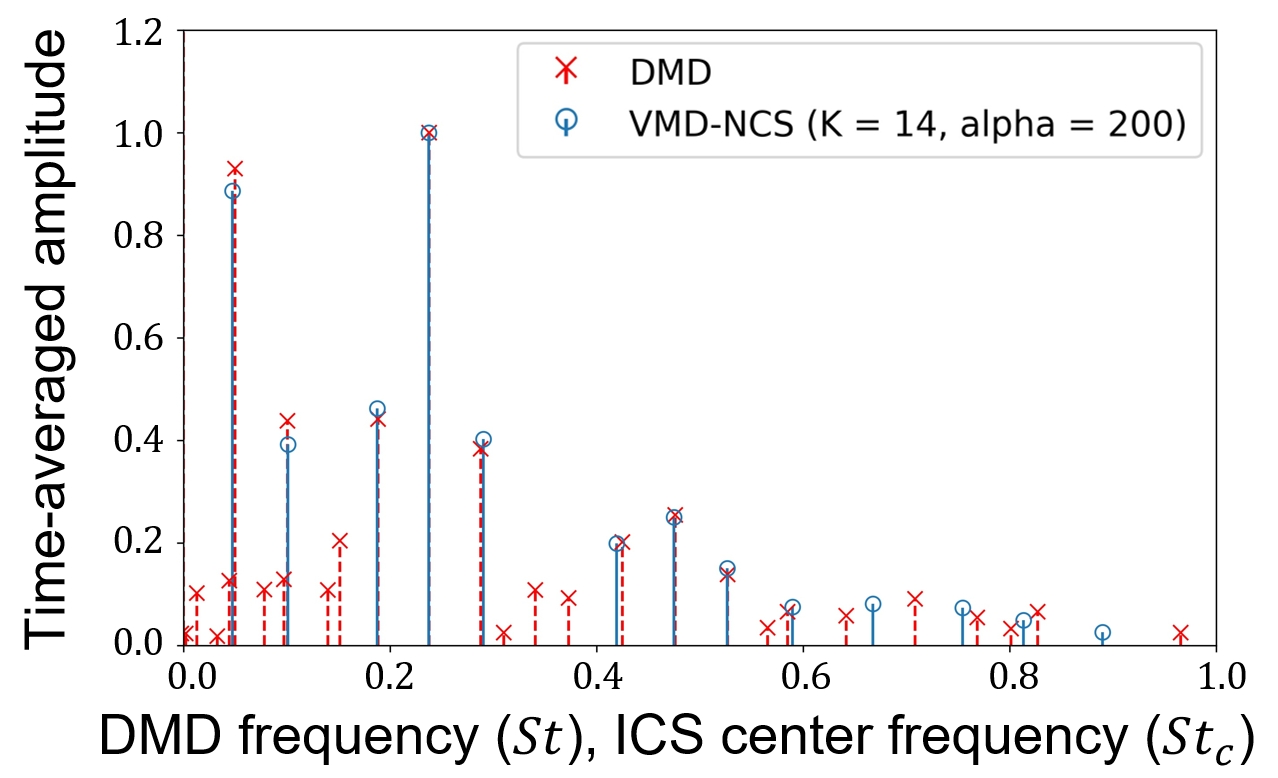}
    \caption{Comparison between DMD and ICS ($K=14,~\alpha=200$) in terms of frequencies and mean amplitudes.}
    \label{fig:pitching_dmd_ics_amp}
\end{figure}
Fig. \ref{fig:pitching_dmd_ics_amp} illustrates the frequencies and mean amplitudes of the DMD modes.
Here, the TLS-DMD algorithm \cite{Hemati2017} was employed for the DMD, and the mean amplitudes of the DMD modes were computed following Kou et al. \cite{Kou2017}.
For comparison, the ICSs for $K=14$ and $\alpha=200$ are also included in the figure.
From this figure, it is evident that the distribution of the center frequencies and mean amplitudes of ICSs for $K=14$ closely resembled those of the DMD modes.
Specifically, the modes of the vortex shedding frequencies plus and minus the pitching frequencies appeared, and their amplitudes closely matched between the ICSs and DMD modes.
This tendency of the ICS obtained at a large $K$ to represent more single-frequency variations (i.e., similar to the case with the DMD modes) was also observed in the analysis described in Section \ref{sec:TWBC}.

In summary, the center frequencies of the ICS obtained from this flow field exhibited similarities to those obtained using the DMD method for large $K$ values.
However, the key advantage of ICS became evident for small $K$ values, and the phenomenon of vortex shedding influenced by wing pitching motion was captured as a single ICS as shown in Sec. \ref{ICSpitching}.

\subsubsection{Comparison of the proper orthogonal decomposition coefficients and multivariate variational mode decomposition modes \rm{($K=4$, $\alpha=200$)}}
\begin{figure}[hbt!]
    \centering
    \includegraphics[width=.99\textwidth]{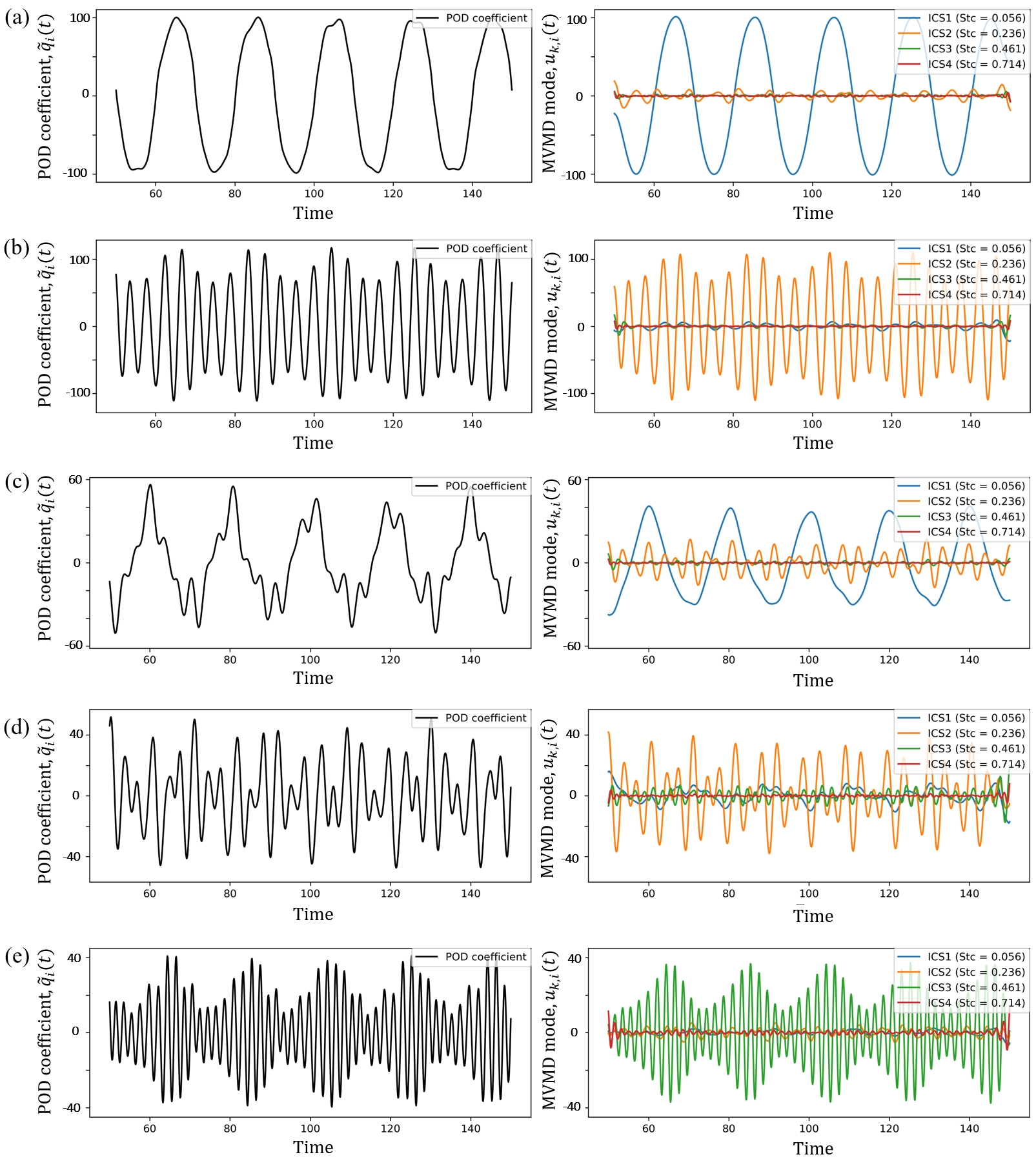}
    \caption{The temporal evolution of the expansion coefficients (left) and MVMD modes obtained by decomposing the expansion coefficients (right) for the (a) first, (b) third, (c) fifth, (d) seventh, and (e) ninth POD modes.}    
    \label{fig:pitching_pod_mvmd_coeff}
\end{figure}

\begin{figure}[hbt!]
    \centering
    \includegraphics[width=.9\textwidth]{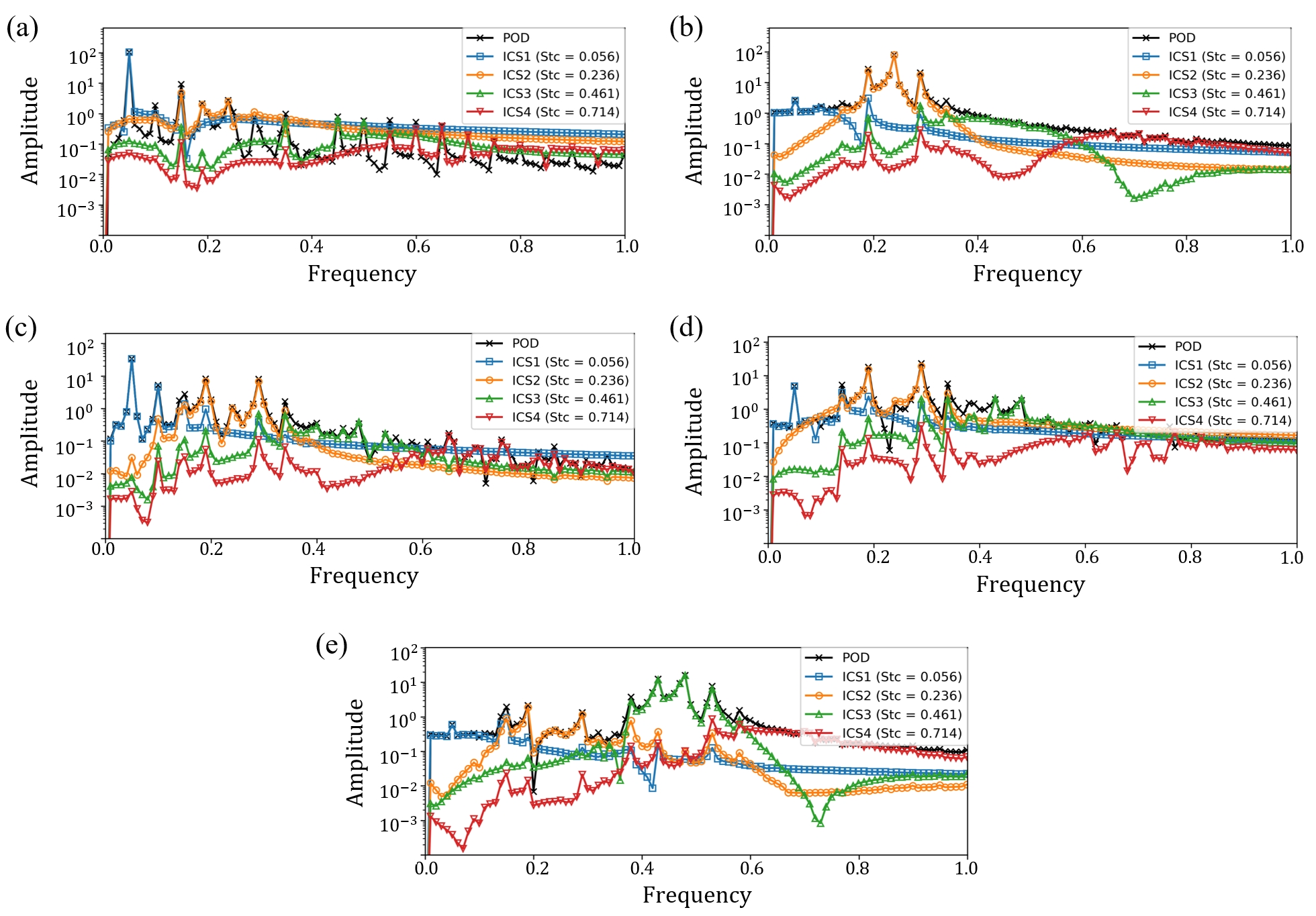}
    \caption{The frequency distribution of the expansion coefficients of the (a) first, (b) third, (c) fifth, (d) seventh, and (e) ninth POD modes and their MVMD modes.}    
    \label{fig:pitching_pod_mvmd_fft}
\end{figure}

Next, the results of the VMD-NCS analysis obtained with $K=4$ and $\alpha=200$ are discussed. 
Fig. \ref{fig:pitching_pod_mvmd_coeff} shows the temporal evolution of the POD expansion coefficients $\tilde{q}_i(t)$ and the MVMD modes $u_{k,i}(t)$. Additionally, Fig. \ref{fig:pitching_pod_mvmd_fft} presents their frequency distributions.
From this figure, similar to the results in Section \ref{sec:TWBC}, it is evident that each POD coefficient exhibits multiple frequency-band fluctuations.
For example, the temporal evolution of the fifth POD mode coefficients showed a combination of variations related to both pitching and vortex shedding frequencies, as seen in Figs. \ref{fig:pitching_pod_mvmd_coeff}(c) and \ref{fig:pitching_pod_mvmd_fft}(c).
Additionally, variations associated with the pitching frequency appeared not only in the fifth but also in the first POD mode. Similarly, variations linked to the vortex shedding frequency were present across multiple modes: the third, fifth, and seventh POD modes.
Therefore, to extract the fluid phenomena associated with each frequency band, it is necessary to isolate and superimpose the frequency components dispersed among multiple POD modes.
The VMD-NCS analysis achieves this by decomposing the POD coefficients using MVMD.
From the frequency distributions (Fig. \ref{fig:pitching_pod_mvmd_fft}), it can be confirmed that the temporal variations of the MVMD modes, which constitute the ICSs, primarily exhibited frequency components around their corresponding center frequencies ($St_c=0.56,~0.236,~0.461,~0.714$).

\subsubsection{Temporal behavior of the intrinsic coherent structure \rm{($K=4$, $\alpha=200$)}}\label{ICSpitching}
\begin{figure}[hbt!]
    \centering
    \includegraphics[width=.5\textwidth]{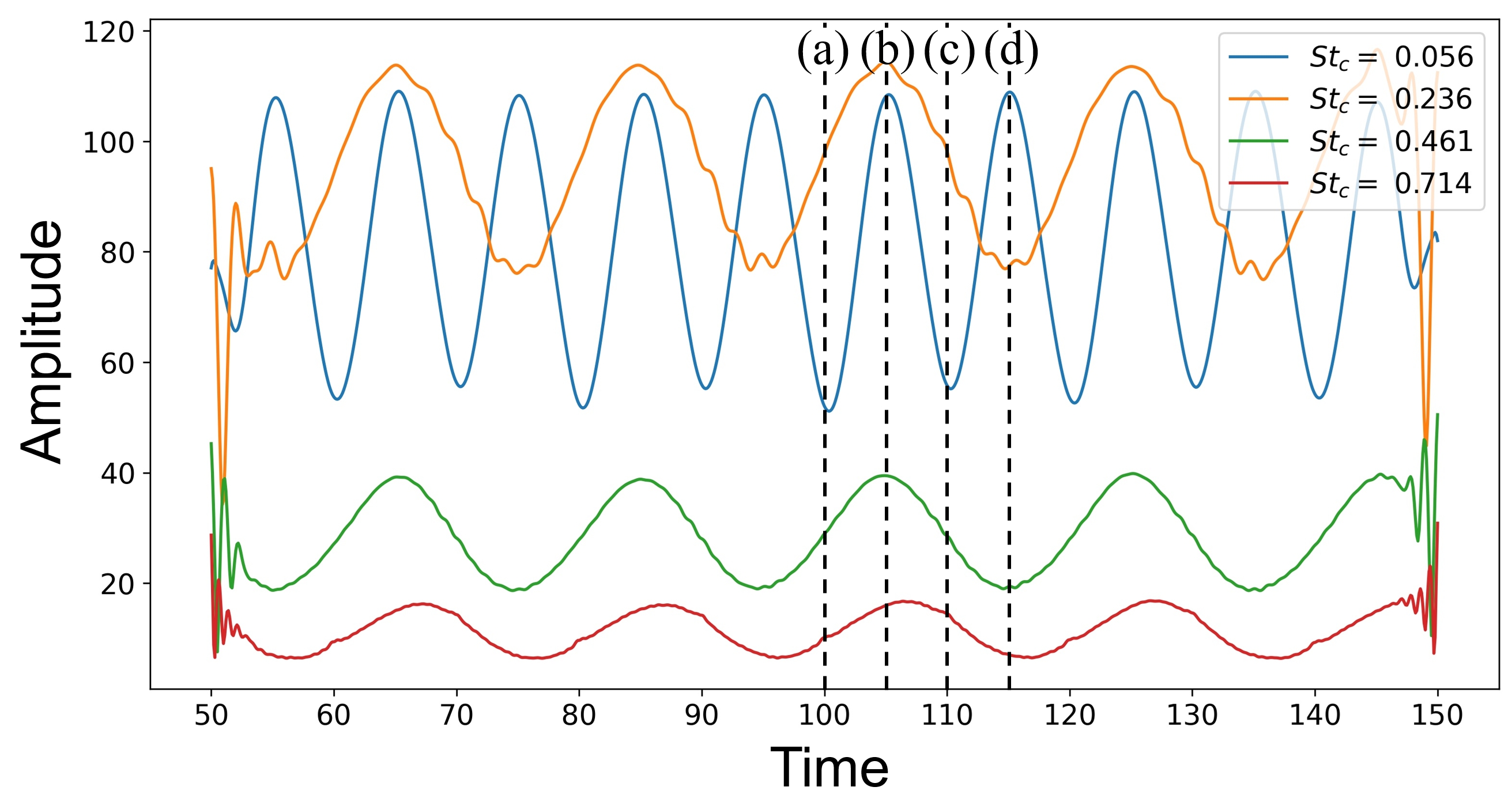}
    \caption{Time history of the ICS amplitudes. (a), (b), (c), and (d) in the figure indicate the times when $a(t)=30^\circ,~25^\circ,~30^\circ$, and $35^\circ$, respectively.}
    \label{fig:pitching_ics_amp_history}
\end{figure}
\begin{figure}[hbt!]
    \centering
    \includegraphics[width=.99\textwidth]{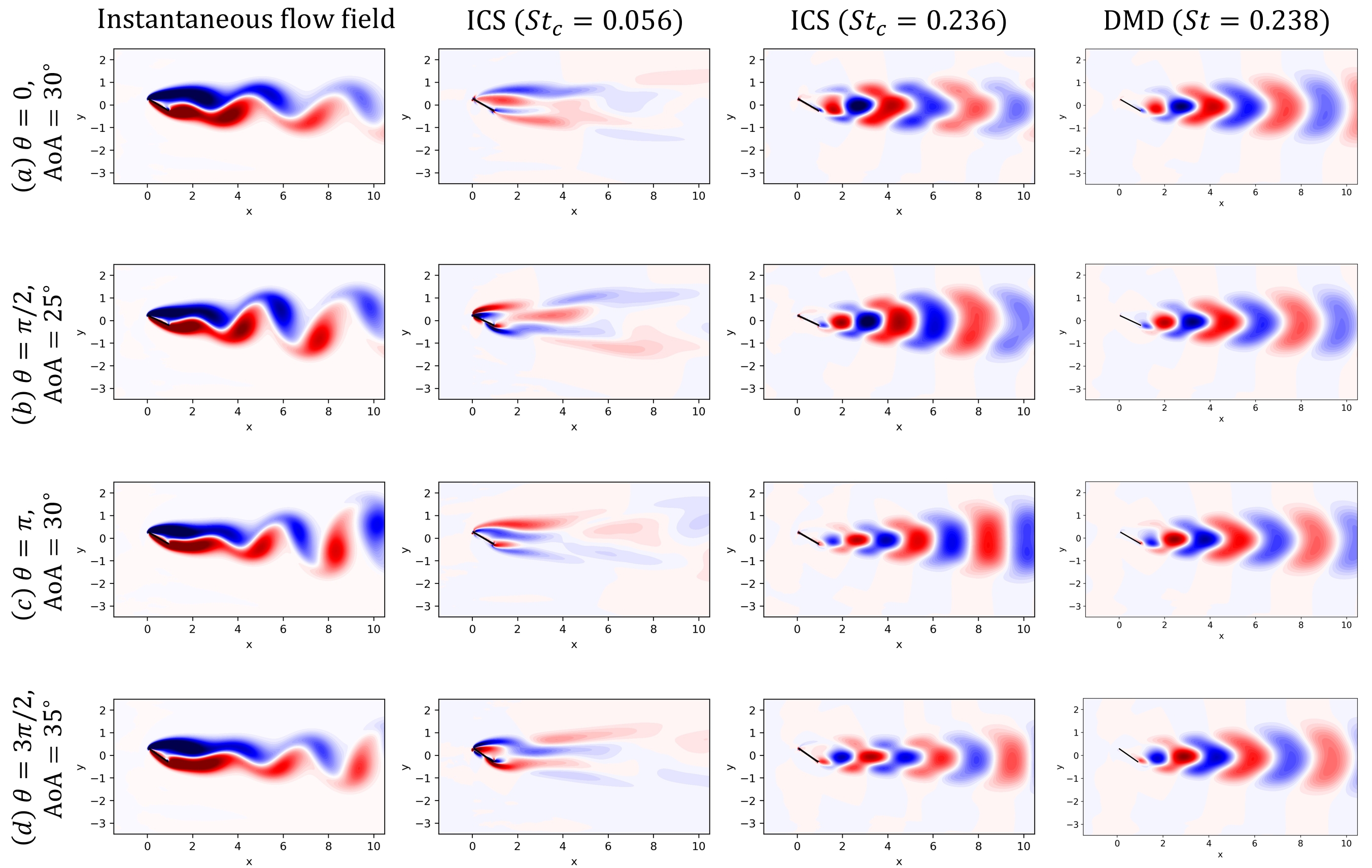}
    \caption{Spatial distribution of the vorticity of the instantaneous flow field, the ICS of $St_c=0.056$, the ICS of $St_c=0.236$, and the DMD mode of $St = 0.238$; (a), (b), (c), and (d) correspond to the times indicated in Fig. \ref{fig:pitching_ics_amp_history}; $\theta$ is the phase in Equation (\ref{eq:phase}), i.e., $\theta = 2 \pi f_p t$.}
    \label{fig:vort_ics_dmd_dist}
\end{figure}

Fig. \ref{fig:pitching_ics_amp_history} shows the time history of the ICS amplitudes, and Fig. \ref{fig:vort_ics_dmd_dist} illustrates the spatial distribution of the instantaneous flow fields and ICSs at each phase (a)--(d) depicted in Fig. \ref{fig:pitching_ics_amp_history}.
By observing the time variation of the amplitude of the ICS corresponding to vortex shedding in Fig. \ref{fig:pitching_ics_amp_history} ($St_c=0.236$), we note that the amplitude reached its maximum at the minimum angle of attack (25 deg) and achieves its minimum at the maximum angle of attack (35 deg).
This indicates that the fluctuations attributed to vortex shedding were influenced by the airfoil's pitching motion, weakened with pitching-up and strengthened with pitching-down.
Conversely, the amplitude of the ICS ($St_c=0.056$) corresponding to the frequency of the pitching motion was not significantly affected by the vortex shedding.

From the spatial distribution shown in Fig. \ref{fig:vort_ics_dmd_dist}, it can be observed that the ICS of $St_c=0.236$ described the influence of the pitching motion on the spatial patterns of the vortex shedding phenomena.
For instance, comparing Figs. \ref{fig:vort_ics_dmd_dist}(a) and (c), at the same angle of attack of 30 deg, it is clear that the spatial pattern of the ICS ($St_c=0.236$) differs, indicating greater fluctuation near the airfoil during pitching-down (a) compared with that during pitching-up (c).

For comparison with the DMD mode, the spatial distributions of the DMD mode ($St = 0.238$) are shown in the fourth column of Fig. \ref{fig:vort_ics_dmd_dist}.
Here, the spatial distribution of each DMD mode at time $t = t_j$, denoted as $\psi_{DMD}(\bm x, t_j)$, is defined as its contribution to the flow field $q(\bm x, t_j)$ when $q(\bm x, t_j)$ is expressed as a superposition of all DMD modes.
It was calculated using the following equation:
\begin{equation}
    \psi_{DMD}(\bm x, t_j) = 2 {\rm{Re}}\left(w(t_j)\phi_{DMD}(\bm x)\right)
\end{equation}
Here, $w(t_j)$ represents the expansion coefficient at time $t = t_j$, $\phi_{DMD}$ represents the DMD eigenvector, and ${\rm Re}(\cdot)$ denotes the real part of a complex number.
As the spatial distribution of the DMD eigenvector, $\phi_{DMD}$, remains unchanged over time, it is evident that the spatial distribution of the DMD modes corresponding to the vortex shedding frequency ($St = 0.238$) was not influenced by the phase of the pitching motion.
As demonstrated in Fig. \ref{fig:vort_ics_dmd_dist}, an advantage of the ICS obtained by the VMD-NCS analysis is its capability to represent spatial-distribution changes in time, which cannot be expressed by a single DMD mode.

\section{Conclusions}\label{Conclusions}
In this study, we developed a data-driven VMD-NCS analysis method for extracting coherent structures from spatiotemporal data containing nonstationary phenomena, such as transient, nonperiodic, or intermittent phenomena, and investigated its effectiveness. 
The VMD-NCS analysis method utilizes low-dimensional representation through POD and nonstationary signal processing through MVMD, and it enables the extraction and analysis of ICSs containing nonstationary phenomena from high-dimensional spatiotemporal data, such as fluid dynamics data.

Validation targeting the transient growth phenomenon around a square cylinder revealed that the VMD-NCS analysis method can extract spatiotemporal patterns (as ICSs) representing not only the periodic shedding of the Kármán vortices but also the transitional phenomena from steady flow to periodic flow.
Additionally, the extraction of ICS depicting the gradual temporal change of the background flow was demonstrated.
A close examination of the temporal variation in the amplitudes of these ICSs revealed observations, for example, an increase in the amplitude of the transient phenomena resulted in  vortex generation along with a decrease in the amplitude of the background flow.
Moreover, during the transient growth phase of vortex shedding, the ICS represented the temporal variation in the spatial distribution, initially showing fluctuations in the far-wake region and gradually intensifying fluctuations in the near-wake region.
In the analysis of the flow around the pitching airfoil, we examined the mechanism through which the VMD-NCS analysis method represents quasi-periodic flow fields resulting from the airfoil's periodic pitching motion and the vortex shedding in the wake region.
The results indicated that the size and amplitude of vortices in the wake region were influenced by the pitching motion, i.e., weakened and strengthened during pitching-up and pitching-down, respectively, represented as temporal changes in a single ICS.
In all these analyses, the ICS derived through VMD-NCS demonstrated a `flexible' attribute \cite{Noack2016}, allowing the capture of temporal changes in spatial patterns as a single ICS, a capability not typically achievable through conventional modal analysis techniques, such as POD and DMD.

Moreover, through investigating the parameter dependence of VMD-NCS analysis, it was found that while parameter $\alpha$, related to the bandwidth, needs to be moderately large to preserve temporal coherence, parameter $K$ exerts a more significant influence on the results of the VMD-NCS analysis.
Increasing $K$ yielded ICSs that represented more periodic variations akin to the results obtained using the DMD method.
For instance, in the pitching airfoil analysis, a relatively high $K$ resulted in ICSs with center frequencies and amplitudes closely matching the DMD modes.
Conversely, reducing the $K$ allowed for frequency changes in ICS with time, capturing a more nonstationary nature.

Thus, the VMD-NCS analysis serves as an effective method for extracting spatiotemporal patterns of nonstationary phenomena, which is difficult to achieve using conventional methods.
The ICS extracted using the VMD-NCS analysis method is `flexible,' evolving with time in terms of spatial distribution and maintaining temporal coherence, thus improving the comprehension of phenomena.
However, the dependency of results on the parameters, i.e., $K$ and $\alpha$, remains a limitation.
Future studies will delve further into understanding the parameter dependence and exploring methods for determining appropriate parameters.

\section*{Acknowledgment}
This work was supported in part by JSPS KAKENHI Grant Number 20K14958, and JST PRESTO Grant Number JPMJPR23O2, Japan.

\bibliographystyle{elsarticle-num}

\bibliography{ref}

\end{document}